\documentstyle[12pt]{article}
\input epsf

\def\0{{\bf 0}}
\def\1{{\bf 1}}
\def\2{{\bf 2}}

\def\NPB#1#2#3{Nucl. Phys. {\bf B#1}, #2 (19#3)}

\def\PL#1#2#3{Phys. Lett. {\bf #1}, #2 (19#3)}

\def\RMPP#1#2#3{Rev. Mod. Phys. {\bf #1}, #2 (20#3)}

\def\bea{\begin{eqnarray}}
\def\eea{\end{eqnarray}}
\def\be{\begin{equation}}
\def\ee{\end{equation}}
\def\nn{\nonumber}

\begin{document}

\title{\textbf{{Generation of large scale magnetic fields by coupling
to curvature and dilaton field}}}
\author{\textbf{ A. Akhtari Zavareh,
A. Hojati and B. Mirza   \footnote{b.mirza@cc.iut.ac.ir} }}
\date{}

 \maketitle {\footnotesize{{\sl{~~~~~~~~~~~~~~~~~Department of Physics,
Isfahan
 University of Technology, Isfahan  84156, Iran}}}}


\begin{abstract}
We investigate the generation of large scale magnetic fields in
the universe from quantum fluctuations produced in the
inflationary stage. By coupling these quantum fluctuations to the
dilaton field and Ricci scalar, we show that the magnetic fields
with the strength observed today can be produced. We consider two
situations: First, the evolution of dilaton ends at the onset of
the reheating stage. Second, the dilaton continues its evolution
after reheating and then decays. In both cases, we come back to
the usual Maxwell equations after inflation and then calculate
present magnetic fields.

\end{abstract}



\vfill


\section{Introduction}

   It is well understood that magnetic fields are present on various
scales in the universe. These scales vary from planet size to
clusters of galaxies of  Mpc order\cite{kronberg}. The problem of
obtaining a reasonable mechanism for generation of such a large
scale magnetic field is an important question in modern cosmology
because of its direct influence on the evolution of universe and
astrophysical events. Magnetic fields have an important role in
the dynamics of galaxies by confining the cosmic rays or
transferring angular momentum away from protostellar clouds so
that they may collapse to become stars (without the loss of
angular momentum, protostellar clouds would collapse to a
low-density centrifugally supported, unstarlike state). They also
play an important role in the dynamics of pulsars, white dwarfs,
and even black holes. The strength of these magnetic fields varies
from $\mu$G in galaxies and cluster of galaxies to a few G in
planets and up to $10^{12}$ G in neutron stars.\\

Some mechanisms known as {\it galactic dynamo} are established to
amplify the scale and strength of magnetic fields by transforming
the kinetic energy of turbulent motion of interstellar medium
into magnetic energy. The gravitational collapse of matter to
form galaxies and cluster of galaxies is definitely an
amplification mechanism that can affect the strength of these
fields because of the magnetic flux conservation. These
mechanisms, however, are only amplification mechanisms and
require a seed magnetic field to amplify to strengths presently
observed . Theories established for generation of these seed
fields are classified into two groups: astrophysical processes
and cosmological processes in the early
universe.\\

If the scale of magnetic fields of galaxies and cluster of
galaxies are in the order of Kpc and Mpc, it means that we should
investigate their origin in the early universe rather than in
 astrophysical processes. Then, these seed magnetic fields are
trapped in highly conductive plasma collapsed to form structures
like galaxies and their clusters during an adiabatic compression
and, finally, subjected to some amplification mechanism like the
 galactic dynamo.\\

Many different mechanisms have been proposed for generation of
these seed magnetic fields in the early universe \cite{D} that may
be
classified as follows :\\

1. Breaking of conformal invariance of electromagnetic
interaction at inflationary stage. This can be realized either
through new non-minimal (and possibly non gauge invariant)
coupling of electromagnetic field to curvature~\cite{turner88},
or in dilaton electrodynamics~\cite{ratra92}, or by the well
known conformal anomaly in the trace of the stress tensor induced
by quantum corrections to Maxwell
electrodynamics~\cite{dolgov93}.\\

2. First order phase transitions in the early universe producing
bubbles of new phase inside the old one~\cite{hogan83}. A
different mechanism but also related to phase transitions is
connected with topological defects, in particular,
cosmic strings~\cite{vachaspati91}.\\

3. Creation of stochastic inhomogeneities in cosmological charge
asymmetry, either electric~\cite{dolgov-s93}, or e.g.
leptonic~\cite{dolgov01}, at large scales which produce turbulent
electric currents and, in turn, magnetic fields.\\

It seems that inflation is the most natural way for overcoming
the large correlation scale \cite{turner88,kolb}. Inflation
produces effects on scales much larger than Hubble horizon in a
natural way. Then, if electromagnetic quantum fluctuations had
been produced in that epoch, they could have been present as large
scale magnetic fields today. The idea is based on the assumption
that a quantum mechanical mode (in scales much smaller than
Hubble Horizon) is excited which freezes when passing through  the
horizon. The problem arises from the fact that conformally
invariant theory can not produce nonmassive particles in a
conformally flat gravitational background. This is what happens to
photons in a FRW background and, therefore, electromagnetic fields
can not be produced. And if the origin of magnetic fields of
galaxies and cluster of galaxies were quantum fluctuations
produced in the inflationary stage, then conformal invariance of
Maxwell theory should have been broken in that
epoch. This happens in several ways \cite{D}.\\

We break the conformal invariance of Maxwell theory by coupling
gravity (Ricci scalar)and a scalar field (dilaton) to it. A
non-minimal coupling of electromagnetic fields to gravity was
introduced first by Turner and Widrow \cite{turner88}. Also, the
coupling of a scalar field to electromagnetic fields has been
studied by different people \cite{turner88,ratra92,B,R}. We mix
these two situations and consider the more realistic condition
that both gravity and a scalar field, that is dilaton field, are
coupled to electromagnetic field during the inflationary era. In
\cite{turner88}, the problem is solved qualitatively and
different values for magnetic fields are derived by changing the
parameters. On the other hand, the problem is treated
parametrically and more quantitatively in \cite{B}. By integrating
these solutions, we get a more generalized situation in which the
coupling parameters are fixed by using some special values for the
magnetic field.\\

We first introduce our theory, then derive equations of motion and
solve them to get an expression for electromagnetic vector
potential, $A_{\mu}$. Then we derive the evolution of electric and
magnetic fields before and after inflation by using joining
conditions. We assume
two situations here :\\

1. Evolution of dilaton field ends by finishing inflation when
dilaton freezes.\\

2. A more realistic situation in which dilaton continues its
evolution
after reheating and then decays into radiation.\\

Finally, we calculate today's strength of magnetic fields on
different scales.
\section{Action and the equations of motion}

For investigating the evolution of magnetic fields produced by a
Maxwell theory whose conformal invariance is broken, we first
introduce the lagrangian. We will then derive the equations of
motion \cite{turner88,ratra92,B}.\\

We introduce inflaton and dilaton scalar fields lagrangian
density as

\bea &&{\cal L}_{dil} =
-\frac{1}{2}g^{\mu\nu}{\partial}_{\mu}{\Phi}{\partial}_{\nu}{\Phi}-
V[\Phi] \;,\label{1-4}\\ [2mm] &&{\cal L}_{inf} =
-\frac{1}{2}g^{\mu\nu}{\partial}_{\mu}{\phi}{\partial}_{\nu}{\phi}
- U[\phi]\;,\label{2-4}\\ [2mm] &&V[\Phi] = \bar{V}
\exp(-\bar{\lambda} \kappa \Phi)\;.\label{6-4}
 \eea
 $U[\phi]$ and $V[\Phi]$ are inflaton and dilaton potentials,
$\bar{\lambda}$ is constant and $\kappa=\frac{8 \pi}{M_{PL}^2}$.
The forms of $U[\phi]$ and $V[\Phi]$ are determined from higher
dimension theories reduced to four dimensions
\cite{D18,D17,R}.\\

The lagrangian density of electromagnetic field coupled to the
dilaton and gravity is \bea &&{\cal L}_{EM} =
f(\Phi)(-\frac{1}{4} F_{\mu\nu}F^{\mu\nu}+\xi RA^2)
\;,\label{3-4}\\
&&F^{\mu\nu} = {\partial}^{\mu}A^{\nu} - {\partial}^{\nu}A^{\mu}
\;,\label{4-4}\\
&&R= 6 \left(\frac{\ddot{a}a+{\dot{a}}^2+k}{a^2}\right)
\;.\label{7-4} \eea $f(\Phi)$ and $\xi RA^2$ are dilaton and
gravitational couplings to electromagnetic field, $k$ is
curvature constant and $R$ is Ricci scalar \cite{ratra92}. $a$ is
scale factor and $\xi$ is a dimensionless parameter which will be
determined. \\

 Therefore, the action is
\be {\cal S} = \int d^{4}x \sqrt{-g} \hspace{3mm} [{\cal
L}_{inf}+ {\cal
L}_{dil}+ {\cal L}_{EM}] \;.\label{8-4}\\
\ee where $g$ is the metric tensor. We consider the metric of
space-time as flat FRW metric (k=0)
 \be
{ds}^2 = g_{\mu\nu}dx^{\mu}dx^{\nu} = {dt}^2 - a^2(t)~
d{\vec{x}}^2
\;.\label{9-4}\\
\ee

Our work is based on four assumptions:\\

1. In the inflationary stage with slow roll condition, the energy
density of inflaton field is much bigger than that of the dilaton
field,
$\rho_{\phi}\gg \rho_{\Phi}$.\\

2. The universe becomes immediately hot after the inflationary
stage,
$t > t_{R}$.\\

3. The conductivity of the universe is ignorable in the
inflationary stage because density of charged particles is very
small in that epoch. After reheating, a lot of charged particles
are produced so that conductivity immediately jumps to a large
value after inflation, $\sigma_c \gg H$.\\

4. We consider two different situations for dilaton. First, the
evolution of dilaton field ends by finishing inflation and the
dilaton freezes; therefore, the coupling will be removed ($f=1$)
\cite{ratra92}. Second, the dilaton continues its evolution after
reheating until it reaches its minimum potential and then decays
into radiation when again $f=1$ \cite{B}.\\

\subsection{Equations of motion}

From action (\ref{8-4}), the inflaton, dilaton, and
electromagnetic potential equations of motion are given by
 \bea
&&-\frac{1}{\sqrt{-g}}{\partial}_{\mu}
(\sqrt{-g}g^{\mu\nu}{\partial}_{\nu}\phi) + \frac{dU[\phi]}{d\phi} = 0\;,\label{10-4}\\
&&-\frac{1}{\sqrt{-g}}{\partial}_{\mu} (
\sqrt{-g}g^{\mu\nu}{\partial}_{\nu}\Phi) + \frac{dV[\Phi]}{d\Phi}
=\frac{d f(\Phi) }{d\Phi} \left( -\frac{1}{4}
F_{\mu\nu}F^{\mu\nu}+\xi
RA^2\right) \;,\label{11-4}\\
&&-\frac{1}{\sqrt{-g}}{\partial}_{\mu} (\sqrt{-g} f(\Phi)
F^{\mu\nu} -2 \xi R A^{\nu}) = 0 \;.\label{12-4}\eea

Since every inhomogeneity of space will be diluted in the
inflationary stage, we can ignore the spatial dependence of the
inflaton and the dilaton scalar fields. The right hand side of
Eq.(\ref{11-4}) is very small and can be considered as a
perturbation in the dilaton
theory.\\

The inflaton and dilaton equations of motion are derived from
Eqs.(\ref{10-4}) and (\ref{11-4})
 \bea
&&\ddot{\phi} + 3H\dot{\phi} + \frac{dU[\phi]}{d\phi} =
0\;,\label{13-4}\\
&&\ddot{\Phi} + 3H\dot{\Phi} + \frac{dV[\Phi]}{d\Phi} =0
\;.\label{14-4} \eea H is the Hubble constant and is derived from
Friedman equation
 \be
H^2 =  \left( \frac{\dot{a}}{a} \right)^2 =
\frac{{\kappa}^2}{3}({\rho}_{\phi} + {\rho}_{\Phi}) \;.\label{15-4}\\
\ee where
 \bea
&&{\rho}_{\phi} = \frac{1}{2}{\dot{\phi}}^2 + U[\phi]\;,\label{16-4}\\
&&{\rho}_{\Phi} = \frac{1}{2}{\dot{\Phi}}^2 +
V[\Phi]\;.\label{17-4} \eea are the inflaton and dilaton energy
densities. Dot implies time derivatives and
$\rho=\rho_{\phi}+\rho_{\Phi}$ is total density. According to our
assumptions, $\rho_{\phi} \gg \rho_{\Phi}$ and we can ignore the
dilaton energy density in the inflationary stage whereby we will
have:

\be
 H^2 \approx \frac{{\kappa}^2}{3} U[\phi] \equiv
{H_{inf}}^2\;.\label{18-4}\\
\ee

We have used slow roll condition in the above equation and
$H_{inf}$ is the value of Hubble constant in the inflationary
stage. Since we assume that the term $ \xi R A^2$ is small
compared to $F_{\mu\nu}F^{\mu\nu}$, we can use the Coulomb gauge,
$A_{0}(t,\vec{x})=0$ and $ \partial_{j}A^{j}(t,\vec{x})=0 $, as
 used in the standard Maxwell theory. We derive equation of
motion for electromagnetic potential $A$ (in comoving coordinate)
from Eq.(\ref{12-4})
 \be
 \ddot{A_i}(t,\vec{x}) +\left( H + \frac{\dot{f}}{f}
\right) \dot{A_i}(t,\vec{x})-\left( \frac{1}{a^2}{\partial}_j
{\partial}_{j}-2 \xi R \right) A_{i}
(t,\vec{x}) = 0\;.\label{19-4}\\
\ee

\subsection{Electromagnetic potential $A$}

To solve the equation of motion for $A$, we first quantize the
theory. The corresponding momentum from ${\cal{L}}_{EM}$ is \bea
&&\pi_{\nu} = \frac{\partial {\cal L}_{EM}}{\partial
{\dot{A_{\nu}}}}\;,\nn \\
&&{\pi}_0 = 0, \hspace{5mm} {\pi}_{i} = f(\Phi)a(t)
\dot{A_i}(t,\vec{x})\;.\label{20-4}\eea

Commutation relation for $A_i$ and $\pi_i$ is \be [\hspace{0.5mm}
A_i(t,\vec{x}), {\pi}_{j}(t,\vec{y}) \hspace{0.5mm} ] = i \int
\frac{d^3 k}{{(2\pi)}^{3}} e^{i \vec{k}
 \cdot ( \vec{x} - \vec{y} )}~~\left( {\delta}_{ij} - \frac{k_i
k_j}{k^2 } \right)\;.\label{21-4}\\
\ee where $k$ is comoving wave number. Using these relations, we
can  write the quantum form of $A$ as \cite{R} \be A_i(t,\vec{x})
= \int \frac{d^3 k}{{(2\pi)}^{3/2}}  [ \hspace{0.5mm}
\hat{b}(\vec{k})A_i(t,\vec{k})e^{i \vec{k} \cdot \vec{x} } +
{\hat{b}}^{\dagger}(\vec{k}) {A_i}^*(t,\vec{k})e^{-i \vec{k} \cdot
\vec{x}} \hspace{0.5mm} ]\;.\label{22-4}\\
\ee in which $\hat{b}$ and $\hat{b^{\dag}}$ are annihilation and
creation operators with the following commutation relations \bea
&&[ \hspace{0.5mm} \hat{b}(\vec{k}),
{\hat{b}}^{\dagger}({\vec{k}}^{\prime})] \hspace{0.5mm} ={\delta}^3 (\vec{k}-{\vec{k}}^{\prime})\;,\nn \\
&&[ \hspace{0.5mm} \hat{b}(\vec{k}),
\hat{b}({\vec{k}}^{\prime})\hspace{0.5mm} ] =[
\hspace{0.5mm}{\hat{b}}^{\dagger} (\vec{k}),
{\hat{b}}^{\dagger}({\vec{k}}^{\prime})]\hspace{0.5mm}  =
0\;.\label{23-4}\eea

For simplicity, we choose $x_1$ along the direction of $\vec{k}$.
Thus, from now on we will work only with two transverse components
(I=2,3). Equation of motion for electromagnetic potential Fourier
modes will be:
 \be
\ddot{A_I}(t,k) + \left( H_{inf} + \frac{\dot{f}}{f}
\right)\dot{A_I}(t,k) +\left( \frac{k^2}{a^2} - 2\xi R\right)
A_{I}(t,k) =
0\;.\label{24-4}\\
\ee
 Ricci scalar in the inflationary stage is $R=12 H^2$ \cite{kolb}.\\

We get normalization condition for $A_i(t,k)$ from Eq.(\ref{21-4})
as \be A_i(t,k){\dot{A}}_j^{*}(t,k) -
{\dot{A}}_j(t,k){A_i}^{*}(t,k) = \frac{i}{fa} \left(
{\delta}_{ij} - \frac{k_i k_j}{k^2
}\right)\;.\label{25-4}\\
\ee
 For more simplicity in solving the equations, we use the following
approximation for the evolution of $f(\Phi)$ \be f(\Phi) =
f[\Phi(t)] = f[ \hspace{0.5mm} \Phi (a(t)) \hspace{0.5mm} ]
\equiv f_{0} a^{2 \alpha-2}\;.\label{26-4}\\
\ee where $f_0$ is a constant and all time variations of $f(\Phi)$
have been put in $a^{2 \alpha -2}$. It can be shown that time
dependence of $\alpha$ can be ignored if we have slow roll
condition for inflaton and dilaton fields \cite{B}. The only
problem that remains is determination of  an acceptable range for
variation of $\alpha$ that can be achieved from consistency
conditions (section 4).\\

From Eq.(\ref{26-4}) \be
H_{inf} + \frac{\dot{f}}{f} = (2 \alpha -1) H_{inf}\;.\label{27-4}\\
\ee By introducing a new variable $d\eta = \frac{dt}{a}$, and
relying on  the fact that $\eta = -\frac{1}{a H_{inf}}$ in the
inflationary stage ( in De'sitter spacetime ), we can rewrite
eq.(\ref{24-4}) as
 \be
\frac{d^2 {A_I}(k,\eta)}{d{\eta}^2}+ \left( \frac{1-2 \alpha}{
\eta }\right) \frac{d {A_I}(k,\eta)}{d \eta} + \left( k^2 + \frac
{24\xi}{{\eta}^2}\right) {A_I}(k,\eta) = 0\;.\label{28-4}\\
\ee One of the forms of Bessel Equation is \be
\frac{d^2{u}}{dz^2} + \left(\frac{ 1 - 2 \nu }{z}\right)
\frac{d{u}}{dz} + \left( \delta^2 + \frac{ {\nu}^2 -
{\nu^{\prime}}^2 }{z^2}\right) u
= 0 \;.\label{29-4}\\
\ee where $\nu$, $\nu'$ and $\delta$ are positive and the solution
is \be u = z^{\nu} ( D_{1} H_{\nu^{\prime}}^{(1)}( \delta z) +
D_{2}
H_{\nu^{\prime}}^{(2)}( \delta z))  \;.\label{30-4}\\
\ee $H_{\nu'}^{(1)}$ and $H_{\nu'}^{(1)}$ are Hankel functions of
first and second order $\nu'$, respectively. Comparing Eqs.
(\ref{28-4}) and (\ref{29-4}), we will have \bea && \eta = z
~~~~~~~~~~~~~~~~~~~ \nu = \alpha ~~~~~~~~~~~~~~~~~~~ \delta
= \pm k \;,\label{31-4}\\
&&\nu^{\prime} =~ \pm \sqrt{{\nu}^2 - 24 \xi }~ =~ \pm
\sqrt{{\alpha}^2 - 24 \xi }\;.\label{32-4} \eea The solution
obtained for Eq.(\ref{28-4}) is
 \be A_I(k,\eta) = D_{1}(a) {(- H_{inf}
\eta)}^{\nu} H_{{\nu}^{\prime}}^{(1)} (-k \eta) + D_{2}(a) (-
H_{inf}
\eta)^{\nu} H_{\nu^{\prime}}^{(2)} (-k \eta)\;.\label{33-4}\\
\ee $D_1$ and $D_2$ coefficients are determined from the
renormalization relation (\ref{25-4}) \be |D_{1}(a)|^2 -
|D_{2}(a)|^2 = \frac{\pi}{4H_{inf} f(a)}~ a^{2 \nu-1}
\;.\label{34-4}\\
\ee For simplicity, we choose \be D_{1}(a) = \sqrt{
\frac{\pi}{4H_{inf}} f(a)} ~~ a^{\nu - 1/2} \hspace{1mm}
e^{i(\nu^{\prime}+1)\pi/4}
\hspace{5mm} D_{2}(a) = 0 \;.\label{35-4}\\
\ee

 Since we are working with large scale
magnetic fields, we expand Eq.(\ref{33-4}) whithin large
wavelength limit for $\nu > 0$ and $\nu < 0$  \bea &&A_I(k,a) =
2^{ \nu^{\prime}} \sqrt{ \frac{1}{4 \pi H_{inf}f(a)} } \Gamma (
\nu^{\prime} ) a^{-1/2} \nn \\
&&~~~~~~~~{\left( \frac{k}{aH_{inf}} \right)}^{-\nu^{\prime}}
e^{i (\nu^{\prime}-1)\pi / 4} ~~~~~~ \nu^{\prime} > 0
\;,\label{39-4-1}\\[4 mm]
&&A_I(k,a) = 2^{- \nu^{\prime}} \sqrt{ \frac{1}{4 \pi
H_{inf}f(a)} } \Gamma (- \nu^{\prime} ) a^{-1/2} \nn \\
&&~~~~~~~~{\left(\frac{k}{aH_{inf}} \right)}^{ \nu^{\prime}} e^{i
(3- \nu^{\prime})\pi / 4} ~~~~~~ \nu^{\prime} < 0 \;.\label{39-4}
\eea

The electric and magnetic fields can be easily derived from the
above expressions using the relation between $\vec{A}$ and
$\vec{E}$ and $\vec{B}$ in comoving coordinate system.
\section{Electric and magnetic fields evolution}

In Coloumb gauge, electric and magnetic fields are derived from
vector potential $A$ as follows
 \be
\vec{E} = - \frac{\partial}{\partial t} \vec{A} , ~~~~~~~~~~~~~
\vec{B}
=\vec{ \nabla} \times \vec{A} \;.\label{40-4}\\
\ee In comoving coordinate, electric and magnetic fields are
\cite{R} \bea
&&{E_i}^{C}(t,\vec{x}) = \dot{A_i}(t,\vec{x})\;,\label{41-4} \\
&&{B_i}^{C}(t,\vec{x}) = \frac{1}{a} {\epsilon}_{ijk} \partial_j
A_k (t, \vec{x})  \;.\label{42-4} \eea where "c" implies comoving
and $\partial_j$ is differentiation in comoving coordinate.\\

Using relations (\ref{39-4-1}), (\ref{41-4}), and (\ref{42-4}), we
can derive Fourier components of electric and magnetic fields in
the inflationary stage

\bea &&{E_I}^{C}(k,a) =  \sqrt{\frac{\pi}{4H_{inf}f(a)}}
\hspace{1.2mm}( \frac{k}{a} ) a^{-1/2} [(\nu - 1/2)
H_{\nu^{\prime}}^{(1)} \left(\frac{k}{a H_{inf}}\right) \nn \\
\hspace{1.2mm} &&~~~~~~~~~~~~~+ 1/2 \left(H_{\nu^{\prime} - 1
}^{(1)} \left( \frac{k}{a H_{inf}}\right) -
H_{\nu^{\prime}+1}^{(1)} \left( \frac{k}{a
H_{inf}}\right) e^{i (\nu^{\prime} + 1) \pi/4}  \right)]\,\label{43-4}\\
&&{B_I}^{C} (k,a) = -i(-1)^I  \sqrt{\frac{\pi}{4H_{inf}f(a)}} \nn
\\ \hspace{1.2mm}
&&~~~~~~~~~~~~~\times \left( \frac{k}{a} \right) a^{-1/2}
H_{\nu^{\prime}}^{(1)} \left( \frac{k}{a
H_{inf}}\right)e^{i(\nu^{\prime}+1)\pi/4} \;.\label{44-4} \eea

 We assumed that the dilaton freezes after
inflation. In this epoch, the conductivity of the universe
increases immediately so that $\sigma \gg H$ \cite{ratra92}. And
the evolution of electromagnetic vector potential follows from the
equation

\be \ddot{A_i}(t,\vec{x}) +\left( \frac{\dot{a}}{a} + {\sigma}_{c}
\right)\dot{A_i}(t,\vec{x})
 - \frac{1}{a^2}{\partial}_j{\partial}_j {A_i}(t,\vec{x}) =
0\;.\label{45-4}\\
\ee

Ratra has shown that we have the following joining conditions for
electric and magnetic fields at transition from the inflationary
stage to the radiation dominated epoch ( $t=t_R$ )
\cite{ratra92,G}, \bea {E^{C}}_{i(RD)}(t_R, \vec{x})
=\exp(-{\sigma}_{c}  t_{R} )
{E^{C}}_{i(INF)}(t_{R}, \vec{x})\;,\label{46-4}\\
B^{C}_{i(RD)} (t_R, \vec{x}) = B^{C}_{i(INF)} (t_R, \vec{x})
\;.\label{47-4} \eea

According to these relations, in a universe with large
conductivity, the electric field accelerates charged particles
but it  reduces exponentially. From Alf'ven theorem, the magnetic
flux is conserved in a conductive universe and, therefore, the
magnetic field evolution is by $a^{-2}$ in the physical
coordinate system \cite{ratra92,R,G}.\\

In the physical coordinate, the electric and magnetic fields are
\cite{R}

\bea &&{E_i}^{Ph}(t,\vec{x})  = a^{-1}E_i^{C}(t,\vec{x})
\;,\label{48-4}\\ [3mm] &&{B_i}^{Ph}(t,\vec{x})  =
a^{-1}B_i^{C}(t,\vec{x}) \;.\label{49-4}\eea Using relations
(\ref{42-4}), (\ref{44-4}) and (\ref{49-4}) we have
 \bea
&&|{B_I}^{Ph}(t,k)|^2
\equiv a^{-2} | {B_I}^{C}(k,a) |^2  \nn \\[5mm]
&&~~~~~~~~=  a^{-2} \left( \frac{\pi}{4 H_{inf} f(a)}
\right){\left( \frac{k}{a} \right)}^2 \left( \frac{1}{a}\right)
H_{\nu^{\prime}}^{(1)} \left( \frac{k}{aH_{inf}} \right)
H_{\nu^{\prime}}^{(2)} \left( \frac{k}{a H_{inf}}
\right)\;.\label{50-4} \eea

We are considering the magnetic fields on large scales. Thus, we
expand relation (\ref{50-4}) for small k to note that the magnetic
field evolves with $a^{-2}$ after inflation \bea
&&|B_I^{Ph}(t,k)|^2  = \frac{ 2^{2 \nu^{\prime} - 2}}{\pi}
\Gamma^2 (\nu^{\prime}) f^{-1}(a_R) \nn \\ [3mm] &&
\hspace{3.5cm} \times\left( \frac{k^2}{a_R H_{inf}}\right) {\left(
\frac{k}{a_R H_{inf}} \right) }^{-2 \nu^{\prime}}
 {\left( \frac{1}{a}\right) }^4, \hspace{10mm}  \hspace{1.4mm}
\nu^{\prime} > 0 \;,\label{51-4}\\[3mm]
&&|B_I^{Ph}(t,k)|^2  = \frac{2^{-2( \nu^{\prime} +1)}}{\pi}
{\Gamma}^2
(- \nu^{\prime}) f^{-1}(a_R) \nn \\[3mm]
&& \hspace{3.5cm} \times   \left( \frac{k^2}{a_R H_{inf}}\right) {
\left( \frac{k}{a_R H_{inf}} \right) }^{2 \nu^{\prime}}
 {\left( \frac{1}{a}\right) }^4, \hspace{10mm}  \hspace{1.4mm}
\nu^{\prime} < 0 \;.\label{52-4}\eea

The magnetic energy density in the Fourier space is \be
{\rho}_B(t,k) =\frac{1}{2}|{B_{t}}^{Ph}(t,k) |^2 ~f(a)\;.\label{53-4}\\
\ee Since two transverse components are equal, $B_t=\sqrt{2} B_I$.
Multiplying (\ref{53-4}) by phase space density $\frac{4 \pi
k^3}{3(2 \pi)^3}$, the large scale magnetic field energy density
in the coordinate space is \be {\rho}_B(L,t) =\frac{k^3}{6
\pi^2}|{B_I}^{Ph}(t,k) |^2
f(a)\;.\label{54-4}\\
\ee
where $L=\frac{2 \pi}{k}$ is correlation length.\\

By integrating Eqs.(\ref{51-4}), (\ref{52-4}) and (\ref{54-4}), we
obtain the magnetic field energy density on the scale
$L=\frac{2\pi}{k}$ \be {\rho}_B(L,t_0) =
\frac{2^{2|\nu^{\prime}|-3}}{3 \pi^3} {\Gamma}^2 (|\nu^{\prime}|)
{H_{inf}}^4 \left( \frac{a_R}{a_0} \right)^4
  { \left( \frac{k}{a_{R} H_{inf}} \right)
}^{-2|\nu^{\prime}|+5}\;.\label{55-4}\\
\ee Accoring to the above relation, the spectrum of the magnetic
fields is invariant for $2|\nu'|=5$.

\section{Consistency condition}

The energy density of electric and magnetic fields should be
smaller than the energy density of dilaton field in the
inflationary stage so that only $V[\Phi]$ determines the evolution
of dilaton \cite{R}. We show the ratio of electromagnetic energy
density to dilaton energy density by $\Theta(L,t)$:
 \bea
 &&\Theta (L, t) \equiv \frac{1}{
{\rho}_{\Phi} }[ \hspace{0.5mm} {\rho}_B (L, t) + {\rho}_E (L, t)
\hspace{0.5mm} ]
\;,\label{56-4}\\
&&{\rho}_E(L,t) = \frac{k^3}{6{\pi}^2} |{E_I}^{Ph}(t,k)|^2
\hspace{0.3mm} f(a)\;.\label{57-4}\eea

If $\Theta \ll 1$, the consistency condition is satisfied. Using
Eqs.(\ref{18-4}), (\ref{43-4}), (\ref{44-4}), (\ref{56-4}) and
(\ref{57-4}), we have \bea &&\Theta \approx \frac{1}{9 w} \left(
\frac{H_{inf}}{M_{Pl}}\right)^2 [ \left(
\frac{k}{aH_{inf}}\right)^5 \times [ H_{\nu^{\prime}}^{(1)}
\left( \frac{k}{a H_{inf}} \right) H_{\nu^{\prime}}^{(2)} \left(
\frac{k}{a H_{inf}} \right) \nn \\ [2.5mm] &&+\left(\left( 1/2
\left(H_{\nu^{\prime} -1 }^{(1)}\left(\frac{k}{a H_{inf}}\right) -
H_{\nu^{\prime}+1}^{(1)}\left(\frac{k}{a H_{inf}}\right)\right) +
\left(\nu - \frac{1}{2}\right) H_{\nu^{\prime}}^{(1)} \left(
\frac{k}{a H_{inf}} \right)\right)\right) \nn
\\ [2.5mm]
&& \times \left( 1/2 \left(H_{\nu^{\prime} -1
}^{(2)}\left(\frac{k}{a H_{inf}}\right) -
H_{\nu^{\prime}+1}^{(2)}\left(\frac{k}{a H_{inf}}\right)\right) +
\left(\nu - \frac{1}{2}\right) H_{\nu^{\prime}}^{(2)} \left(
\frac{k}{a H_{inf}} \right)\right)]] \nn
\\
[2.5mm] \;.\label{60-4}\eea

As a measure of $\Theta$, one may use the upper limit of $H_{inf}$
calculated from CMB anisotropy observations \cite{B15,B26,B24}
 \be
\frac{H_{inf}}{M_{Pl}} \leq 2 \times 10^{-5}
~~~~~~~\Longrightarrow
~~~~~~~~  H_{inf}= 2.4 \times 10^{14} GeV \;.\label{61-4}\\
\ee and noting that $\frac{k}{aH_{inf}}$ decreases with evolution
of universe, from Eq.(\ref{60-4}), we may conclude that when
$\frac{k}{aH_{inf}} = 1$, $\Theta < 10^{-10} \omega^{-1}$ (
$\omega$ is defined as
$\omega\equiv\frac{V[\Phi]}{\rho_{\phi}}\approx\frac{V[\Phi]}{U[\phi]}$
and according to our assumption that the dilaton energy density is
much smaller than that of inflaton in the inflationary era,
$\omega \ll 1$ ), the consistency condition exists. By
investigating these relations within large wavelength limits, we
have
 \bea
 &&\Theta \approx \frac{1}{9 \pi^2 w} {\left(
\frac{ H_{inf}}{M_{Pl}}\right)}^2  [ x^{- 5 + 2 \nu^{\prime}} 2^{2
\nu^{\prime}} \Gamma^2 ( \nu^{\prime} )+ 1/4 ( 2^{ 2(
\nu^{\prime} - 1 )} \Gamma^2 (\nu^{\prime} -1 ) x^{-2} \nn \\
[2.5mm] && -  2^{2 \nu^{\prime} + 1 } \Gamma (\nu^{\prime} -1 )
\Gamma ( \nu^{\prime} + 1 ) +  2^{(2 \nu^{\prime} + 1) } \Gamma^2
( \nu^{\prime} + 1 )
 x^{2} )+ \nn \\[2.5mm]
&& ( \nu - 1/2)^2 2^{2 \nu^{\prime} } \Gamma^2 (\nu^{\prime} ) +
( \nu -1/2 )(- 2^{2 (\nu^{\prime} + 1) } \Gamma (\nu^{\prime} +1
) \Gamma ( \nu^{\prime} )
 x \nn \\[2.5mm]
 &&+ 2^{2 \nu^{\prime} } \Gamma (\nu^{\prime} ) \Gamma ( \nu^{\prime} -
1 ) x^{-1})] ]
~~~~~~~~~~~~~~~~~~~~~~~~~~~~~~~~~~~~~~~~~~~~~~\nu^{\prime}
> 1 \;,\label{62-4}\\ [3mm] &&\Theta \approx \frac{1}{9 \pi^2 w}
{\left( \frac{ H_{inf}}{M_{Pl}}\right)}^2  [ x^{- 5 + 2
\nu^{\prime}} 2^{2 \nu^{\prime}} \Gamma^2 ( \nu^{\prime} )+ 1/4 (
2^{ 2(1 - \nu^{\prime})} \Gamma^2 (1 - \nu^{\prime} ) x^{2(1 - 2
\nu^{\prime})} \nn \\ [2.5mm] && -  2^{2} \Gamma (\nu^{\prime} +1
) \Gamma (1 - \nu^{\prime} ) x^{2(1 -  \nu^{\prime}} + 2^{(2
\nu^{\prime} + 1) }  \Gamma^2 ( \nu^{\prime} + 1 )
 x^{2} )+ ( \nu - 1/2)^2 2^{2 \nu^{\prime} } \Gamma^2 (\nu^{\prime} )
\nn \\[2.5mm]
&& + ( \nu -1/2 ) (2^{2} \Gamma (\nu^{\prime} ) \Gamma (1 -
\nu^{\prime} )
 x^{1 - 2  \nu^{\prime}} - 2^{2 \nu^{\prime} + 1} \Gamma (\nu^{\prime}
) \Gamma ( \nu^{\prime} + 1 ) x)] \nn \\[2.5mm]
&&
~~~~~~~~~~~~~~~~~~~~~~~~~~~~~~~~~~~~~~~~~~~~~~~~~~~~~~~~~~~~~~~~~~~~~~~~~~~~0
< \nu^{\prime} < 1  \;,\label{63-4}\\ [3mm] &&\Theta \approx
\frac{1}{9 \pi^2 w} {\left( \frac{ H_{inf}}{M_{Pl}}\right)}^2  [
x^{-( 5 + 2 \nu^{\prime})} 2^{- 2 \nu^{\prime}} \Gamma^2 ( -
\nu^{\prime} )+ 1/4 ( 2^{ 2(1 - \nu^{\prime})} \Gamma^2 (1 -
\nu^{\prime} ) x^{2} \nn \\ [2.5mm] && -  2^{2} \Gamma
(\nu^{\prime} +1 ) \Gamma (1 - \nu^{\prime} ) x^{2} + 2^{(2
\nu^{\prime} + 1) }  \Gamma^2 ( \nu^{\prime} + 1 )
 x^{2(1 + \nu^{\prime})} )+ ( \nu - 1/2)^2 2^{- 2 \nu^{\prime} }
\Gamma^2 (- \nu^{\prime} ) \nn \\[2.5mm]
&& + ( \nu -1/2 ) ( - 2^{2} \Gamma (- \nu^{\prime} ) \Gamma (1 +
\nu^{\prime} )
 x^{1 + 2  \nu^{\prime}} + 2^{2 (1 - \nu^{\prime} )} \Gamma (-
\nu^{\prime} ) \Gamma (1 - \nu^{\prime} ) x)] \nn \\[2.5mm]
&&~~~~~~~~~~~~~~~~~~~~~~~~~~~~~~~~~~~~~~~~~~~~~~~~~
~~~~~~~~~~~~~~~~~~~~~~~~~~-1 < \nu^{\prime} < 0
 \;,\label{64-4}\\ [3mm]
&&\Theta \approx \frac{1}{9 \pi^2 w} {\left( \frac{
H_{inf}}{M_{Pl}}\right)}^2  [ x^{-( 5 + 2 \nu^{\prime})} 2^{- 2
\nu^{\prime}} \Gamma^2 (- \nu^{\prime} )+ 1/4 ( 2^{ 2(-
\nu^{\prime} - 1 )} \Gamma^2 (-\nu^{\prime} -1 ) x^{-2} \nn \\
[2.5mm] && -  2^{-2 \nu^{\prime} + 1 } \Gamma (-\nu^{\prime} -1 )
\Gamma ( -\nu^{\prime} + 1 ) +  2^{(-2 \nu^{\prime} + 1) }
\Gamma^2 ( -\nu^{\prime} + 1 )
 x^{2} )+ ( \nu - 1/2)^2 2^{-2 \nu^{\prime} } \Gamma^2 (-\nu^{\prime} )
\nn \\[2.5mm]
&& + ( \nu -1/2 )(- 2^{2 (-\nu^{\prime} + 1) } \Gamma
(-\nu^{\prime} +1 ) \Gamma ( -\nu^{\prime} )
 x + 2^{-2 \nu^{\prime} } \Gamma (-\nu^{\prime} ) \Gamma (
-\nu^{\prime} - 1 ) x^{-1})] \nn \\[2.5mm]
&&~~~~~~~~~~~~~~~~~~~~~~~~~~~~~~~~~~~~~~~~~~~~~~~~~~~~~~~~~~~~~~~~~~~~~~~~~~~~
\nu^{\prime} < -1  \;,\label{65-4} \eea which we defined as
$x=\frac{aH_{inf}}{k}$. Since $\frac{k}{aH_{inf}}$ decreases with
the expansion of the universe, a negative sign must be selected
for the power of  $x$ so that the consistency condition ($ \Theta
\ll 1 $) is maintained. The intervals $0< \nu' <1$ and $-1< \nu' <
0 $ give us trivial results. Therefore, $\nu' >1$ and $\nu' < -1 $
conditions will determine the range of $\nu' $ by considering the
consistency condition. From $\nu' >1$ and $\nu' <-1$, we conclude
$2 \nu' <3 $ and $2\nu'
>-3$, respectively. By integrating the two situations, the
consistency condition implies the range $-3< 2\nu' <3$ for
$\nu'$. Note that the first terms in Eqs.(\ref{62-4})-(\ref{65-4})
is related to the magnetic field and the other terms are related
to the electric field.

\section{Present magnetic field}

As explained before, the seed magnetic fields produced in the
early universe have been affected by amplification mechanisms to
get the strengths and scales observed today. These amplification
mechanisms are {\it galactic dynamo} and {\it gravitational
collapse}.\\

The galactic dynamo mechanism is based on the conversion of the
kinetic energy associated with the differential rotation of
galaxies into the magnetic field energy \cite{D17,t5,t6}. In the
ideal situation, this mechanism could have amplified the magnetic
field strength with a factor of $e^{30}$ or $10^{13}$ during the
$30$ revolutions of protogalaxies since their formation up until
now \cite{D17}. In the real situation, the dynamo mechanism could
have amplified the magnetic field with a factor of $10^7$
although it has not been properly obtained in the cluster of
galaxies.\\

The gravitational collapse is another means for amplifying the
magnetic fields. For the galaxies, right before its formation, a
patch of matter of roughly $1$ Mpc scale collapses by
gravitational instability. Right before the collapse, the mean
energy density of the patch stored in matter is of the order of
the critical density of the Universe. Right after collapse the
mean matter density of the protogalaxy is, approximately, six
orders of magnitude larger than the critical density. Since the
physical size of the patch decreases from $1$ Mpc to $30$ Kpc, the
magnetic field increases because of the flux conservation, of a
factor $(\rho_{\rm a}/\rho_{\rm b})^{2/3} \sim 10^{4}$ where
$\rho_{\rm a}$ and $\rho_{\rm b}$ are, respectively, the energy
densities right after and right before the  gravitational collapse
\cite{D17}. For the cluster of galaxies, their density is greater
than the critical density of the universe by a factor of $10^3$
and we have, $(\rho_{\rm a}/\rho_{\rm
b})^{2/3} \sim 10^{2}$.\\

The presently observed strengths are $10^{-6}$ G in galaxies and
$10^{-7}$ in cluster of galaxies. By both mechanisms available,
one has to have the magnetic fields with strengths of $10^{-23}$ G
(in ideal galactic dynamo) and $10^{-17}$ (in the more realistic
galactic dynamo) for galaxies on $1$ Mpc scale. When just the
gravitational collapse acts on them, we should have the seed
magnetic fields with strengths of $10^{-10}$ G in galaxies and
$10^{-9}$ G in cluster of galaxies (on $10$ Mpc scale).\\

By noting that in Eq.(\ref{55-4}), none of the amplification
mechanisms has been considered for the current magnetic fields, we
consider the values mentioned in the above paragraph as the
present strengths of the magnetic fields of galaxy and cluster of
galaxies and emphasize that the observed fields can be achieved
by the effect of the amplification mechanisms.\\

\begin{figure}
 \centerline{ \epsfxsize=12cm\epsfysize=8.5cm\epsffile{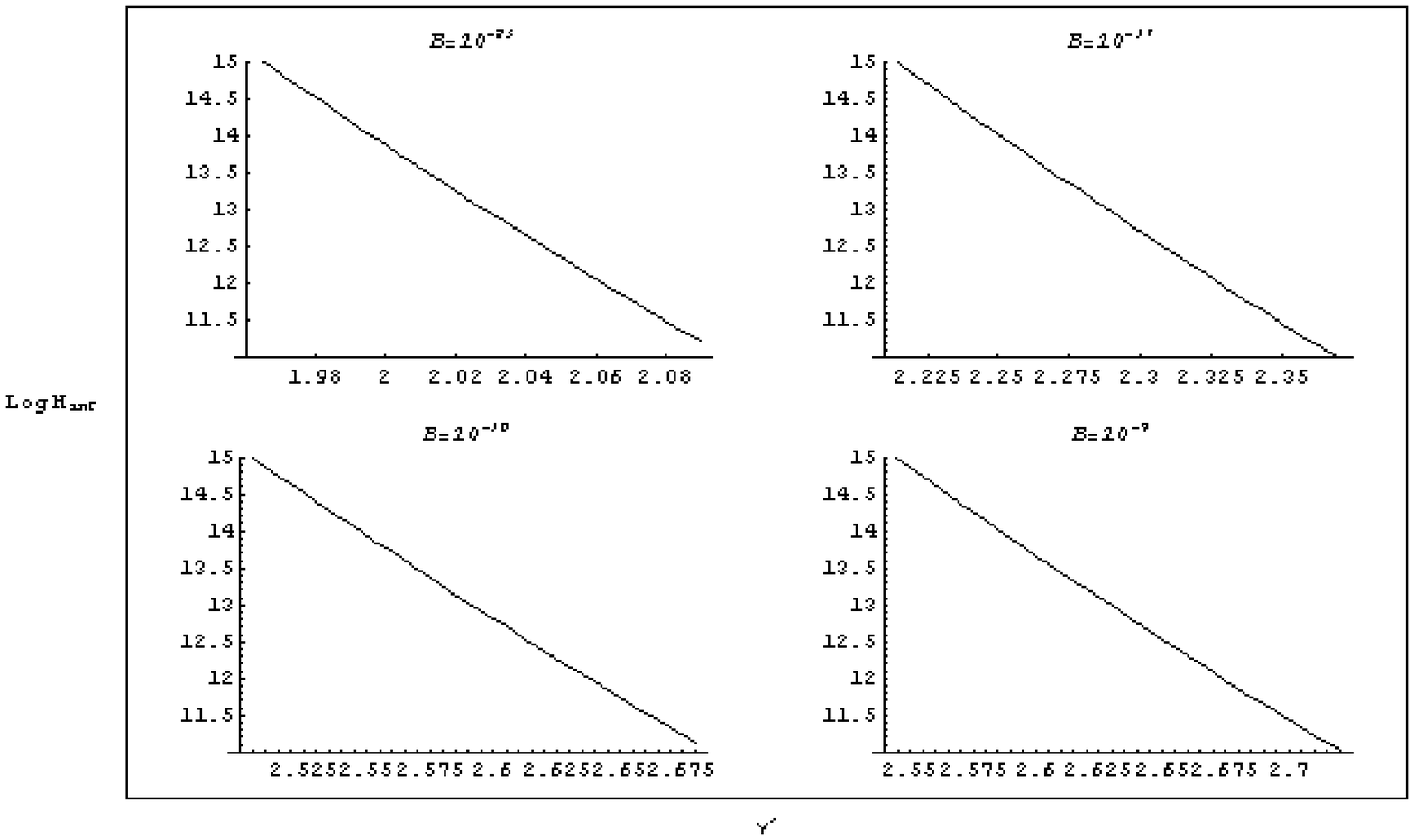}}
 \caption{$Log(H_{inf})$ relative to $\nu^{\prime}$, for magnetic
fields with strengths $10^{-23}$, $10^{-17}$, $10^{-10}$G on 1
MPc scale and $10^{-9}$G on 10 MPc scale, at present.}
\label{fig1}
 \end{figure}

We get the value of magnetic field at the present time as
 \be
{\rho}_B (L, t_0) \propto |B(L, t_0)|^2 \propto \left(
\sqrt{H_{inf} M_{Pl}} L \right)^{|2 \nu^{\prime}|-5} \left( ~~~
T_{ \gamma_0} \sqrt{ \frac{H_{inf}}{M_{Pl}} }~~\right)~~^4
\;.\label{66-4}\\
\ee where we have used the following relations \cite{kolb} \bea
&& H = 1.66 \times g_{*}^{1/2}~~~ \frac{T^2}{M_{PL}}\;,\label{68-4}\\
&&{\rho}_{\phi} = \frac{{\pi}^2}{30} g_{*} {T_R}^4
\hspace{1.7mm}\hspace{5mm}(g_{*} \approx 200)\;,\label{69-4}\\
&&N = 45 + \ln ( \frac{L}{[Mpc]} +\ln { \frac{ [ 30/({\pi}^2
g_{*} )]^{1/12}{{\rho}_{\phi}}^{1/4}
}{10^{38/3}[GeV]}}\;,\label{70-4}\\ [3mm] &&\frac{a_0}{a_{R}} =(
\frac{ g_{*} }{3.91})^{1/3}\frac{ T_{R} }{T_{ \gamma 0}
}\hspace{1mm} \approx \hspace{1mm} \frac{ 3.7 T_{R}}{2.35 \times
10^{-13}[GeV]} \;.\label{71-4}\eea
 $"0"$ implies quantities at present. $T_R$ and
$T_{\gamma_{0}}$ are temperatures of CMB  at the reheating epoch
and at the present time, respectively ( $T_{\gamma_{0}} \approx
2.73 K$ ). $N$ is e-folding between time of first crossing,
$t_1$, and reheating. We can rewrite eq.(\ref{66-4}) as an
equality

 \bea
 &&\frac{1}{2}
\times 4.8 \times 10^{- 39 }~ B(L, t_0)^2 = \frac{2^{2
|\nu^{\prime}|-3}}{3 \pi^2}~ \Gamma^2 (|\nu^{\prime}|)
{\left(\frac{
g_{*} }{3.91}\right)}^{-1/3} \times 1.66 ~~ T_{ \gamma_0} (GeV) \nn \\
[4.5mm] &&   \times ~{g_{*}}^{1/2} (~{\sqrt{
\frac{H_{inf}}{M_{Pl} }}~)}^4 \times {( ~(\frac{(
\frac{30}{{\pi}^2 g_{*}})^{-1/6}} {10^{38/3}[GeV][Mpc]}
\sqrt{H_{inf} M_{Pl}} L~[Mpc] )\times e^{45})}^{|2
\nu^{\prime}|-5}
 \;.\label{67-4}\eea
To find a range for $\nu'$ variations, the graph of
$\log(H_{inf})$ with respect to $|\nu'|$ is plotted
(fig.~\ref{Fig1}). By taking the logarithm of both sides, one has
 \bea
&&Log
(H_{inf})={(\frac{2~\nu^{\prime}-1}{2})}^{-1}(-(16.041(2~\nu^{\prime}-5)\nn \\
&&+2~Log(\Gamma(\nu^{\prime}) +0.6 ~\nu^{\prime})+51.98-2
Log|\vec{B}|)\;.\label{73-4} \eea
 As mentioned before, the upper limit for $H_{inf}$ is
$2.4 \times 10^{-14} GeV $. We can estimate a lower limit for
$H_{inf}$ by noting that duration of the inflationary stage had
been about $10^{-34}$ \cite{ROOS} and that the least value of
e-folding number should be about $60$ \be \frac{a_{i}}{a_{R}} =
e^N = e^{\Delta t~ H} ~~~~~~~\Rightarrow~~~~~~~ H_{inf} = (\Delta
t)^{-1} N = 4 \times 10^{11} GeV \;.\label{74-4}\\
\ee where "i" means value of quantities at the beginning of
inflation. Now, we use Eq.(\ref{73-4}) to depict the graph of
$\log(H_{inf})$ versus $|\nu'|$, for the magnetic fields with
strengths $10^{-23}$, $10^{-17}$,  and $10^{-10}$ G on $1$ Mpc
scale (galaxies) and $10^{-9}$ G  on $10$ Mpc scale (cluster of
galaxies). From Fig.~(\ref{fig1}), we find that for the magnetic
fields mentioned, with $10^{11} < H_{inf} < 10^{15}$, the ranges
of $\nu'$ variations are $1.96 < \nu' <2.9$, $2.21< \nu' <2.37$,
$2.5< \nu' <2.68$ and $2.54< \nu' < 2.75$, respectively. All of
these ranges violate the consistency condition, $|\nu'| <
\frac{3}{2}$. From Eqs.(\ref{62-4})-(\ref{65-4}), it can be seen
that the consistency condition comes from $H_{\nu' +1}^{(1)}
(\frac{k}{a H_{inf}})$ (which is part of the electric field). For
solving this problem, we have to cancel out the term $H_{\nu'
+1}^{(1)}$. Imposing the condition $\nu-\frac{1}{2}=\nu'$ on
eq.(\ref{43-4}) and considering the following relations between
derivations of Hankel functions
 \bea
&&H_{\nu^{\prime}-1}^{(1)} (x) + H_{\nu^{\prime}+1}^{(1)} (x)= 2
\frac{\nu^{\prime}}{x}  H_{\nu^{\prime}}^{(1)}(x) \nn \\
&&H_{\nu^{\prime}-1}^{(1)} (x) - H_{\nu^{\prime}+1}^{(1)} (x)= 2
\frac{d}{dx} H_{\nu^{\prime}}^{(1)}(x) \;.\label{76-4} \eea the
electric field can be written as \be {E_I}^{C}(k,a)
=\sqrt{\frac{\pi}{4H_{inf}f(a)}} \hspace{1.2mm}\left( \frac{k}{a}
\right) a^{-1/2} \left( \frac{k}{a H_{inf}} \right)
H_{\nu^{\prime} - 1 }^{(1)}
 e^{i(\nu^{\prime}+1)  \pi/4}  \;.\label{77-4}
\ee Now rewriting eqs.(\ref{62-4})-(\ref{65-4}) using the above
relation, we arrive at \bea &&\Theta \approx \frac{1}{9 \pi^2 w}
{\left( \frac{ H_{inf}}{M_{Pl}}\right)}^2  x^{- 5 + 2
\nu^{\prime}} (2^{2 \nu^{\prime}} \Gamma^2 ( \nu^{\prime} )+
 2^{ 2( \nu^{\prime} - 1 )} \Gamma^2 (\nu^{\prime} -1 ) x^{-2})
~~~~~~~~~~~~~\nu^{\prime} > 1
 \;,\label{84-4}\\ [2.5mm]
&&\Theta \approx \frac{1}{9 \pi^2 w} {\left( \frac{
H_{inf}}{M_{Pl}}\right)}^2   x^{- 5 + 2 \nu^{\prime}} (2^{2
\nu^{\prime}} \Gamma^2 ( \nu^{\prime} )+
 2^{ 2( \nu^{\prime} - 1 )} \Gamma^2 (\nu^{\prime} -1 ) x^{-2})
 ~~~~~~~~0 < \nu^{\prime} < 1
\;,\label{85-4}\\[2.5mm]
&&\Theta \approx \frac{1}{9 \pi^2 w} {\left( \frac{
H_{inf}}{M_{Pl}}\right)}^2   x^{-( 5 + 2 \nu^{\prime})} (2^{- 2
\nu^{\prime}} \Gamma^2 (- \nu^{\prime} )+
 2^{- 2( \nu^{\prime} + 1 )} \Gamma^2 (-\nu^{\prime} -1 ) x^{-2})
 ~~~ \nu^{\prime} < -1 \;.\label{86-4} \eea

The new consistency condition is $\frac{-3}{2} < \nu^{\prime} <
\frac{5}{2}$. The problem of satisfying the consistency condition
still remains for $\nu' > \frac{-3}{2}$ and, therefore, the
acceptable range for $\nu'$ variation is $\nu' < \frac{5}{2} $.
 The new consistency condition is derived by the redefinition of
 $\nu - \frac{1}{2}=\nu^{\prime}$ while we have $\alpha=\nu$ from
 Eq.(\ref{32-4}) which results in a new consistency condition as
 $\alpha <3$. Using $\nu' = \alpha -\frac{1}{2}$ in Eqs.(\ref{64-4})
and (\ref{73-4}) one has \bea &&\frac{1}{2} \times 4.8 \times
10^{- 39 }~ B(L, t_0)^2 = \frac{2^{2 \alpha-4}}{3 \pi^2}~ \Gamma^2
(\alpha- 1/2) {(\frac{ g_{*} }{3.91})}^{-1/3} \times 1.66 ~~ T_{
\gamma_0} (GeV) \nn \\ [4.5mm] &&   \times ~{g_{*}}^{1/2}
(~{\sqrt{ \frac{H_{inf}}{M_{Pl}}}~})^4 \times {( ~(\frac{(
\frac{30}{{\pi}^2 g_{*}})^{-1/6}} {10^{38/3}[GeV][Mpc]}
\sqrt{H_{inf} M_{Pl}} L~[Mpc] )\times e^{45})}^{2
\alpha-6} \;.\label{87-4}\\[3mm]
&&Log(
H_{inf})={(\alpha-1)}^{-1}(-(16.041(2~\alpha-6)+2~Log(\Gamma(\alpha-
1/2)) \nn \\ [3mm] &&~~~~~~~~~~+0.6 (~\alpha-1/2))+51.98-2
Log|\vec{B}|)\;.\label{88-4} \eea

\begin{figure}
 \centerline{\epsfxsize=12cm\epsfysize=8.5cm\epsffile{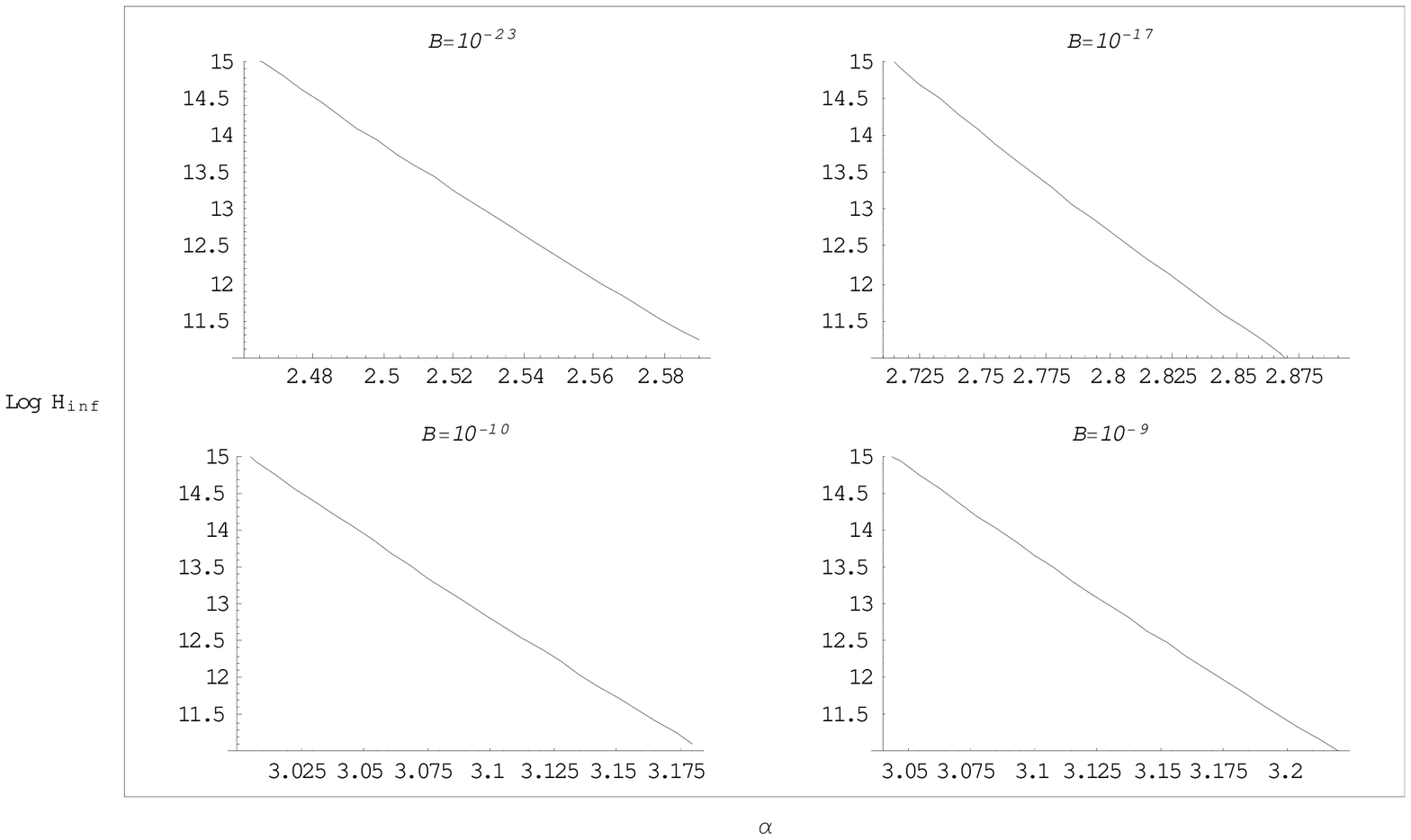}}
 \caption{$Log(H_{inf})$ relative to $\alpha$, for magnetic fields with
strengths $10^{-23}$, $10^{-17}$, $10^{-10}$G on 1 MPc scale and
$10^{-9}$G on 10 MPc scale, at present} \label{fig2}
 \end{figure}

We conclude from Eq.(\ref{88-4}) that the larger $H_{inf}$ and
$\alpha$, the larger will the magnetic field be. For $\alpha=1$,
magnetic field is $2.1 \times 10^{-58}$ G in $1$ Mpc scale which
is independent of $H_{inf}$. For $\alpha=3$, magnetic field
spectrum is invariant, and the maximum value of the field is $1.8
\times 10^{-11}$ G for $H_{inf}=2.4 \times 10^{14} GeV$. By using
Eq.(\ref{88-4}) to depict the graph of $\log(H_{inf})$ versus
$|\nu'|$, for the magnetic fields with strengths $10^{-23}$,
$10^{-17}$  and $10^{-10}$ G on $1$ Mpc scale (galaxies) and
$10^{-9}$ G on $10$ Mpc scale (cluster of galaxies),
Fig.~(\ref{fig2}), it is clear that for
 $10^{-23}$ G and $10^{-17}$ G fields, the consistency condition holds
 in the appropriate range for $H_{inf}$ while it does not for
$10^{-10}$ G and $10^{-9}$ G fields. This should not concern us
 since we got $10^{-10}$ G and $10^{-9}$ G fields by assuming that
 the galactic dynamo does not act on galactic scales which is not,
indeed, realistic.\\

As mentioned before, the most real situation for galaxies is a
$10^{-17}$G magnetic field on $1$ Mpc scale, which transforms to a
$10^{-6}$G field observed today via gravitational collapse and
dynamo action. In this way, a range for $\alpha$ parameter can be
obtained
 \be
2.71 <\alpha<2.8\;.\label{89-4}\\
\ee

Therefore, when dilaton freezes after inflation, the best range
for the coupling $f(\Phi)$ can be determined from
eq.(\ref{89-4}). Considering $f=f_0 a^{2 \alpha -2 }$,
$a=c~t^{1/2}$ (c is a constant) , $f(t_{R})=1$ and $t_{R} \approx
\frac{1}{2H_{inf}}$, we have \be f_0 = c^*~ (2
H_{inf})^{\alpha-1}\;. \nn \ee

Obtaining the value for $f_0$ and having the range of $\alpha$,
 we can determine the dilaton coupling in this case.\\

 From Eq.(\ref{32-4}) and the consistency condition that leads to
$\nu^{\prime}=\nu -1/2=\alpha-1/2$, we get  \be \xi=\frac{\alpha
-1/4}{24} \;.\ee and from Eq.(\ref{77-4}), we have $\xi \simeq
\frac{1.23}{12} $ that is close to the value $\frac{1}{12} $, as
derived by Turner and Widrow in \cite{turner88}. Since $\alpha$
and $\xi$ parameters are specified, the coupling to
electromagnetic field for generating the $10^{-17} G$ magnetic
field is determined.\\

Conventionally, $\xi$ is a constant parameter \cite{turner88} and
the term $\xi R A^2$ will become very small after inflation since
Ricci scalar in FRW metric is: $R=\frac{{\ddot{a}}a +
{\dot{a}}^2+K}{a^2}$ and it will be very small in the radiation
 and matter dominated epoch, hence it will be removed.\\

\section{Dilaton decay}

By now, we have assumed that the dilaton evolution ends after
inflation. Let us now consider a more realistic situation (like
 in \cite{B}) in which the dilaton continues its evolution and
  reaches its minimum  potential, then it begins to oscillate and then
decays (the dilaton evolves from $\Phi_R(< 0)$ to its minimum in
$\Phi=0$ and here, $f=1$).

\subsection{Dilaton evolution after inflationary stage}

During coherent oscillation, the energy density $\rho_{\Phi}$, of
the dilaton field evolves as $a^{-3}$ which is slower than the
energy density of radiation produced by the inflaton field via its
decay, $\rho_{\phi}$. If the condition $\rho_{\Phi} < \rho_{r}$
holds until dilaton decays, the entropy per comoving volume
remains approximately unchanged, but if it changes to $\rho_{\Phi}
> \rho_{r}$, a large amount of entropy will be produced \cite{kolb}. If
this happens, the magnetic energy density will be diluted by
entropy
production.\\

We consider $t_{osc}\simeq m^{-1}$ ($m$ is dilaton mass) as the
time when coherent oscillations begin. When $t>t_R$, the universe
is radiation dominated and thus, $a(t)=a_R (t/t_R)^{1/2}$. Before
coherent oscillation, $t_R<t<t_{osc}$,
the dilaton field evolves with its exponential potential.\\

The amplitude of the dilaton field, $\Phi_{R}$, is relatively
small at the end of inflation. This is a consequence of the
condition we imposed on the magnetic field energy density
(consistency condition). When $t_R<t<t_{osc}$, the magnetic
energy density increases by the dilaton evolution. The magnetic
energy density should be smaller than the dilaton one on all
scales so that it does not affect the dilaton evolution. As will
be shown, larger values of $\bar{\lambda} \kappa |\Phi_R|$ result
in  larger values of the magnetic energy density and, thus, we get
an upper limit for $\bar{\lambda} \kappa |\Phi_R|$ from here. By
solving the dilaton equation of motion with exponential potential
(eq.(\ref{11-4})) numerically, we get a range for $\bar{\lambda}
\kappa |\Phi_R|$ that satisfies these conditions \cite{B}. The
result is that $|\Phi_R|$ with
$\bar{\lambda} \sim {\cal O}(1)$ should be of the order $1/\kappa$. \\

With these conditions and the fact that
$\bar{V}\exp(-\bar{\lambda}\kappa \Phi_R)/\rho_{\phi}$
$\approx\omega \ll 1$, the dilaton energy density will not be
larger than the radiation energy density and we will, therefore,
have entropy production only in the coherent oscillation epoch, $t
\sim t_{osc}$. By considering the slow roll condition, we have
$\rho_{\Phi} \approx V[\Phi]$ for the dilaton field. The dilaton
gets its minimum of potential at $t_{osc}$ and thus, $\rho_{\Phi}
\approx \bar{V}$. The minimum arises when other contributions
from gaugino condensation enter into the to dilaton potential and
generate a minimum \cite{B}. If the universe stays
radiation-dominated while the dilaton decays, $t_{d} \simeq
\Gamma_{\Phi}^{-1}$, then

 \be {{\rho}_{r}}^{(inf)} (t_{\Phi}) >
{\rho}_{\Phi}(t_{\Phi}) \ee or \be {\rho}_{\phi} [
\frac{a(t_{\Phi})}{a_R} ]^{-4} > {\rho}_{\Phi}(t_{osc}) [
\frac{a(t_{\Phi})}{a(t_{osc})}]^{-3} \ee Since, $\rho_{\Phi}
\approx \bar{V}$, one gets
 \be
\frac{{\rho}_{\phi}}{\bar{V}} > \left( \frac{M_{Pl}}{m}
\right)\left( \frac{2H_{inf}}{m}\right)^2\;.\label{81} \ee where,
we have used $\Gamma_{\Phi} \simeq m(m/M_{pl})^2$ and
$t_{R}\approx \frac{1}{2}H_{inf}$. Entropy per comoving volume
remains constant here \cite{kolb}.Thus, the necessary condition
for entropy production is \be \frac{{\rho}_{\phi}}{\bar{V}} <
\left( \frac{M_{Pl}}{m} \right) \left( \frac{2H_{{inf}}}{m}
\right)^2 \;.\label{82} \ee Suppose the dilaton energy density
equals radiation energy density at $t=t_c$ \be
 {\rho}_{\phi}  [\frac{a(t_{c})}{a_{R}} ]^{-4}
  =    {\rho}_{\Phi}(t_{{osc}})
     [ \frac{a(t_{c})}{a(t_{osc})} ]^{-3}
\;.\label{83} \ee We, therefore, have \be
 t_{c} \approx( \frac{{\rho}_{\phi}}{ \bar{V}} )^2  {t_{R}}^4
\hspace{0.2mm}
 {t_{{osc}}}^{-3} \approx  \left( \frac{{\rho}_{\phi}}{ \bar{V} }
\right)^2 \left( \frac{1}{2H_{{inf}}} \right)^4 m^3 \;.\label{84}
\ee where $\rho_{\Phi}(t_{osc}) \approx \bar{V}$ and
$t_{osc}\simeq$
  $m^{-1}$ are assumed. After $t_c$,
 $\rho_{\Phi}$ dominates over $\rho_R$ and
universe becomes matter-dominated and thus
$a(t)=a_c(t/t_c)^{2/3}=a_R t_R^{-\frac{1}{2}} t_c ^{-\frac{1}{6}}
t^{\frac{2}{3}} $ that second equality comes from continuity
condition of $a(t)$ at $t=t_R$.\\

The entropy per comoving volume is $S=a^3 (\rho + p)/T$ where
$\rho$, $p$ and T are energy density, pressure, and temperature,
respectively, in equilibrium \cite{kolb} and we have \bea S^{4/3}
= {S_{c}}^{4/3} + \hspace{1mm}\frac{4}{3}{\rho}_{\Phi}(t_{c})
{a_{c}}^{4} [
 \frac{2{\pi}^2 \langle g_{*}\rangle}{45}]^{1/3} {\Gamma}_{\Phi}
\int_{t_{c}}^{t} [ \frac{a(\tau)}{a_{c}} ]\exp [\hspace{0.5mm} -
{\Gamma}_{\Phi} (\tau - t_{{c}})
 \hspace{0.5mm}   ] d\tau \;,\label{85}\nn \\[4 mm]
{S_{c}}^{4/3} [\hspace{0.5mm} 1+{\Gamma}_{\Phi} {t_{c}}^{-2/3}
\int_{0}^{\infty}(u+t_{{c}})^{2/3} \exp ( - {\Gamma}_{\Phi} u  )
d u \hspace{0.5mm} ] \approx {S_{c}}^{4/3} [ \hspace{0.5mm}
  1+\Gamma ( \frac{5}{3} ) ( t_{{c}} {\Gamma}_{\Phi})^{-2/3}
\hspace{0.5mm} ] \;.\label{86} \eea where $S_c$ is the entropy
per comoving volume at $t_c$ and $<g_{*}>$ is the average of
$g_{*}$ (number of degrees of freedom ) for decay duration. In
the second approximation, we have $u\equiv \tau -t_c$ and we
consider the limit
 $u \rightarrow \infty$. In addition, we have used
$\rho_r(t_c)=\rho_{\Phi}(t_c)$ and the following equation
 \be
 S = [\hspace{1mm}
\frac{4}{3} (\frac{2{\pi}^2 g_{*}}{45} )^{1/3} a^4 {\rho}_{{r}}
\hspace{1mm} ]^{3/4}\;.\label{87} \ee that in general, gives the
relation between the radiation energy density and the entropy per
comoving volume. From $\Gamma_{\Phi} \simeq m(m/M_{pl})^2$ and
Eqs.(\ref{84}) and (\ref{86}), we find that the ratio of entropy
per comoving volume after decay to that of before decay is
written as:

 \be
\Delta S \equiv \frac{S}{S_{c}} \nonumber \\[2mm]
  \approx \{ \hspace{1mm}
   1+ \Gamma ( \frac{5}{3} )  [
 \left(  \frac{\bar{V}}{{\rho}_{\phi}}  \right)
 \left( \frac{2H_{{inf}}}{m} \right)^2
 \left( \frac{M_{Pl}}{m} \right) ]^{4/3} \hspace{1mm}
\}^{3/4}  \nonumber \\[3mm]
\approx
 \left( \frac{\bar{V}}{{\rho}_{\phi}}  \right)
 \left( \frac{2H_{{inf}}}{m} \right)^2
 \left( \frac{M_{Pl}}{m} \right)
\;.\label{88} \ee

By entropy production, the universe expands more rapidly to cancel
this entropy production. It follows from this observation and
also from the relation $\rho_r \propto a^{-4}S^{4/3}$, that \be
 ( \Delta S  )^{4/3} =
\left( \frac{{a_0}^{\prime}}{a_0} \right)^{4}\;.\label{89} \ee
where ${a_0}^{\prime}$ is the present scale factor when we have
entropy
production.\\

Finally, we investigate the effect of dilaton decay on the energy
density of present large scale magnetic fields. Again, we assume
that the universe immediately becomes highly conductive after
reheating. From Eqs.(\ref{55-4}),(\ref{88}), and (\ref{89}), we
find the
 ratio of magnetic field energy density at the present time, when
dilaton continues evolving after reheating, $\rho^{\prime}$, to
the situation that dilaton freezes in the reheating, $\rho$, is
\be
  \frac{\rho^{\prime}}{\rho}
= f^{-1}(t_{R}) ( \Delta S  )^{-4/3}\nn\;,\label{90}\\
\ee We define \be \omega=\frac{V[\Phi]}{\rho_{\phi}} \ee Since we
assumed $\rho_{\phi}>> \rho_{\Phi}$, then $\omega <<1$ which we
consider as $\omega \approx 10^{-2}$.\\

From Eq.(\ref{90}) and $\frac{\bar{V}}{\rho_{\phi}} \approx \omega
exp~(\bar{\lambda} \kappa \Phi_R)$, we arrive at
 \be
 \frac{\rho^{\prime}}{\rho} \approx f^{-1}(t_R)\exp [
\hspace{0.5mm} \frac{4}{3}~ \bar{{\lambda}} \kappa {\Phi}_{R}
\hspace{0.5mm} ][ \hspace{1mm} w (\frac{2H_{{inf}}}{m} )^2 (
\frac{M_{Pl}}{m} ) \hspace{1mm} ]^{-4/3}\;.\label{92}
 \ee

As can be seen from RHS of the above equation, it is the dilaton
coupling that increases $\rho^{\prime}$ relative to $\rho$.

\subsection{Magnetic fields with dilaton decay}

Considering $f=f_0 a^{2 \alpha -2 }$, $a=c~t^{1/2}$ (c is a
constant), $f(t_{osc})=1$ and $t_{osc} \approx \frac{1}{m}$, we
have \be f_0 = c^*~ m^{\alpha-1}\;. \nn \ee

since $t_R=\frac{1}{2 H_{inf}}$, \be
f(t_R)=m^{\alpha-1}~({2H_{inf}})^{1-\alpha}\;. \label{103} \ee

\begin{figure}
\centerline{\epsfxsize=20cm\epsfysize=20cm\epsffile{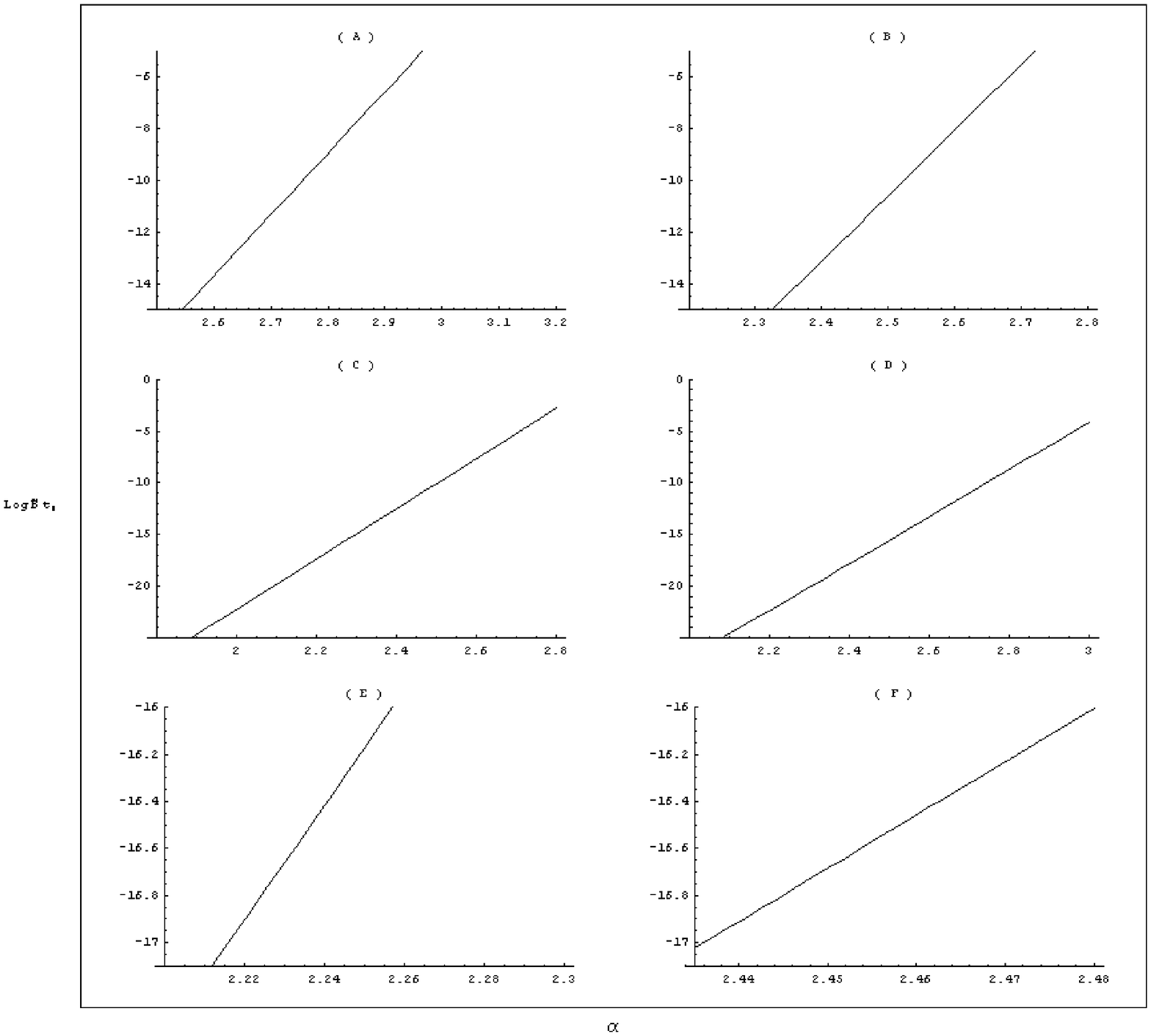}}
 \caption{$Log(B^{\prime}(T_0))$ relative to $\alpha$, for $H_{inf}=2.4
\times 10^{14} GeV$, $m=2.4 \times 10^{13}
 GeV$ in "A", "C" and "E",
 for $H_{inf}= 10^{11} $GeV, $m=10^{10}
$GeV in "B", "D" and "F". Graphs "A" and "B" are on 10 Mpc and
others are on 1 Mpc. Graphs "E" and "F" are depicted to get the
accurate value of $\alpha$. } \label{fig3}
\end{figure}

We estimate from Eqs.(\ref{55-4}), (\ref{92}) and (\ref{103}), the
present strength of magnetic fields for the case in which dilaton
continues evolving after reheating  which leads to the following
relation: \bea Log(
B^{\prime}(t_{0}))=\frac{1}{2}[{(\alpha-1)}Log(H_{inf}) +
c(2~\alpha-6)+2~Log(\Gamma(\alpha- 1/2) \nn \\
+0.6 (~\alpha-1/2))-23.96 + \bar{\lambda} \kappa |\Phi_R|
+(\alpha -1)(Log (2 H_{inf})-Log (m))+ \Delta S ]\;.\label{93}
\eea

c is equal to $16.041$ and $17.041$ on $1$ Mpc and $10$ Mpc scale,
respectively. It can be seen that a  stronger magnetic field
results from a larger value of $\bar{\lambda} \kappa |\Phi_R|$. As
said before, $\bar{\lambda} \kappa |\Phi_R|$ is of the order $1$
and any bigger value violates the consistency condition.
Therefore, we use the maximum possible value for $\bar{\lambda}
\kappa |\Phi_R|(=1)$ and plot the graph for $Log( B^{\prime})$
versus $\alpha$ (fig(\ref{fig3})). The graphs of magnetic field
strengths are on $10$ Mpc scale in "A" and "B" and on $1$ Mpc
scale in "C" and "D".  We have entropy production in all the
cases. The entropy production in "B", "D", and "F" is $4.4 \times
10^9$
while it is about $2 \times 10^5$ in "A", "C", and "E".\\

Comparing these graphs with those of  Fig(\ref{fig2}), one can
find that in this case  smaller values of $\alpha$ give the
desired magnetic fields $( 10^{-23}, 10^{-17}, 10^{-10}~ and
~10^{-9} G)$ and we don't have consistency condition violation in
any of the field strengths.\\

The graphs "E" and "F" represent a strength of $10^{-17}$ (which
is the best choice ) with
$H_{inf}=10^{14}~Gev$ and $H_{inf}=10^{11}~ Gev$, respectively.\\

From graphs "E" and "F", the best range for $\alpha$ to give the
acceptable strength is $ 2.215 < \alpha < 2.441 $ and when  the
value of $\xi$ is obtained from Eq.(\ref{77-4}), the couplings are
completely determined.

\section{conclusion}

In this work, a mechanism is introduced for amplifying quantum
fluctuations in the inflationary universe and for producing large
scale magnetic fields. This is accomplished by breaking the
conformal invariance of electromagnetic theory in the early
universe by coupling the gravity (curvature of spacetime) and a
scalar field (dilaton) to it. These couplings have been studied
separately, before. Here, the more realistic problem to the
effect that both of these
couplings exist together is considered.\\

Two different situations have been discussed for the evolution of
the dilaton scalar field. In the first situation, whereby the
dilaton is assumed to freeze at the end of inflation, the
parameters of the model are determined for the seed magnetic
field that gives the presently observed strengths by the effect of
amplification mechanisms. By considering the strength of
$10^{-17}$ gauss as the best value for the seed magnetic field to
give presently observed field, we fixed our coupling parameters as
: $2.71<\alpha<2.8$ (we use "2.71" that corresponds to the upper
limit of $H_{inf}$ which is more realistic) and $\xi \approx
\frac{1.23}{12}$.\\

In the second situation, which assumes that the dilaton continues
its evolution after the inflation, a large amount of entropy can
be produced that dilutes the energy density of the magnetic
fields produced. Here, we expect less values for $\alpha$ since
our coupling nears unity in a larger time than the previous case.
The range for variations of the parameters that give the desired
magnetic field ( $10^{-17}$ gauss ) includes $2.215<\alpha<2.441$
and $\xi \approx \frac{0.98}{12}$ which is
closer to the value derived in \cite{turner88}.\\

\vskip 0.2in

\noindent {\bf Acknowledgements} \vskip 0.1in

\noindent We would like to thank Prof. B.Ratra, Dr. K.Bamba and
Dr. N.Afshordi for their useful discussion and suggestions and
Isfahan University of Technology for their financial support.

\newpage


\end{document}